# Electric Field Controlled Magnetization and Charge-Ordering in $Pr_{0.6}Ca_{0.4}MnO_3$

Himanshu Sharma[1,a)], Geetha Balakrishnan[2], Don McKenzie Paul[2], A. Tulapurkar[3,b)], C. V. Tomy[1,c)]

[1]*Department of Physics, Indian Institute of Technology Bombay, Powai, Mumbai – 400 076, India.*

[2]*Department of Physics, University of Warwick, Coventry, CV4 7AL, United Kingdom.*

[3]*Department of Electrical Engineering, Indian Institute of Technology Bombay, Powai, Mumbai – 400 076, India.*

In this paper, we present the observation of the electric field control on the charge-ordering and metamagnetic transitions during the magnetization measurements in a single crystal of $Pr_{0.6}Ca_{0.4}MnO_3$ (PCMO). We have demonstrated that the complete melting of charge ordering can be realized in a single crystal of PCMO by applying a voltage as small as 2.5 V, which otherwise needs magnetic fields in excess of 11 T. The maximum change in magnetization with applied voltage occurs across the charge-ordering transition temperature ($T_{CO}$ = ~ 235 K). Even though the electric field does not seem to affect the magnetic ordering, we see a clear evidence at low temperatures for the occurrence of the metamagnetic transitions at higher fields with the application of electric field.

The capability to externally control the properties of magnetic materials is an idea which drives extreme research on magnetic semiconductors and multiferroics[1]. Over the last decade, the external electric field control properties of manganites (e.g., electrical control of magnetic anisotropy, domain structure, spin polarization or critical temperatures[2,3,4]) has attracted much attention for its low-power spintronics[2] and magnetoelectronics[2,3] device applications. In mixed valent manganites, $Pr_{1-x}Ca_xMnO_3$ (For $0.3 \leq x \leq 0.7$) a charge-ordered insulator (COI) phase exists, which can be melted into a metallic phase by extrinsic strain i.e., the application of external electric[5] fields or magnetic[6] fields, electromagnetic radiation[7], pressure[8] etc. Tremendous efforts have been devoted to lower the melting magnetic field for charge ordering in $Pr_{1-x}Ca_xMnO_3$ (PCMO)[6,9,10]. Charge ordering (CO) in manganite is interesting as it competes with double exchange responsible for magnetic ordering and induce numerous interesting properties[9,12]. It is found that the charge ordering of PCMO can be destroyed by applying a large magnetic field as up to 40 T[6]. However, it is found that the melting fields can be lowered in PCMO thin films[11]. Extreme research has been going on in manganite thin films to demonstrate electric field effect on magnetic anisotropy[14], charge-ordered states[5,12,13], colossal magnetoresistance (CMR)[15] and electroresistance (ER)[15] using a ferroelectric or dielectric gate[16-19]. However, in previous researches, mainly transport properties have been studied, but the systematics in term of magnetization were not studied nor a mechanism determined. In this paper, we present the effect of voltage or electric field on the charge ordering (CO) transition during magnetization measurement in a $Pr_{0.6}Ca_{0.4}MnO_3$ (PCMO) single crystal. For a comparison, the effect of magnetic field on the CO transition in the same PCMO single crystal is also investigated.

Single crystals of $Pr_{0.6}Ca_{0.4}MnO_3$ (PCMO) provided by the Warwick group[12], were grown using an infrared image furnace by the floating zone method. A rectangular piece (lateral size 1 mm × 2 mm and thickness of 0.5 mm) of the PCMO single crystal was cut from the as grown rod for the present measurements. Top and bottom faces of the sample (across the thickness) were covered with silver pads for electrical connection (see inset of Fig. 1). Magnetization measurements were carried out using a SQUID Magnetometer (MPMS-XL, Quantum Design Inc). The sample rod was suitably modified to apply the required voltages across the sample from a Keithley Source Meter (Keithley-2602A). The magnetization measurements were recorded as a function of temperature, magnetic field and applied voltage (electric field) across the sample.

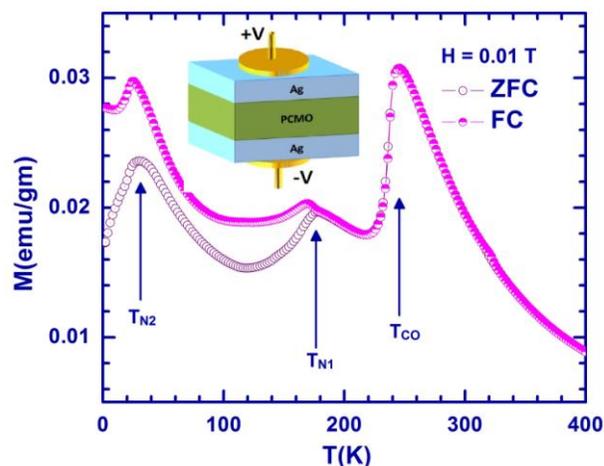

FIG. 1. Zero Field-Cooled (ZFC) and Field-Cooled (FC) magnetization as a function of temperature (T) in an applied magnetic field of 0.01 T for the $Pr_{0.6}Ca_{0.4}MnO_3$ (PCMO) single crystal. Arrows highlight the three transitions, $T_{N1}$, $T_{N2}$ and $T_{CO}$. Inset shows the schematic of sample configuration used for

[a)]*Email: himsharma@iitb.ac.in*  [b)]*Email: ashwin@ee.iitb.ac.in*

[c)]*Email: tomy@iitb.ac.in*

In order to make sure that the attachment of the voltage leads does not affect the magnetization measurements, we measured the zero field-cooled (ZFC) and field-cooled (FC) magnetization of $Pr_{0.6}Ca_{0.4}MnO_3$ (PCMO) single crystal as a function of temperature in an applied magnetic field of 100 Oe with zero applied voltage across the sample. Magnetization of PCMO, as shown in Fig. 1, is almost identical to the reported magnetization[10] with a Charge-Ordering (CO) transition (e.g., $T_{CO}$ = ~235 K) followed by two antiferromagnetic transitions (e.g., $T_{N1}$ ~ 175 K, and $T_{N2}$ ~ 25 K).

To study the effect of electric field on magnetic/charge ordering transitions, we measured the field-cooled magnetization as a function of temperature in an applied magnetic field of 0.5 T after applying different voltages across the sample, the results of which are shown in Fig. 2. Two interesting features are clearly visible in the magnetization data; (i) the charge ordering temperature ($T_{CO}$) shifts drastically (~ 10 K) even for a very small voltage of 0.5 V applied across the sample and (ii) the magnetization decreases with applied voltage in the temperature range, 300 K down to the charge ordering transition temperature. This shifting of the charge-ordering temperature as a function of applied voltage suddenly ceases and the charge ordering disappears altogether for an applied voltage of 2.5 V. Thus, it is clear that the melting of charge ordering occurs for applied voltages ≥ 2.5 V. It is observed that the $T_{CO}$ decreases linearly at a rate of 20 K/V with increase in voltage as shown in the inset of Fig. 2.

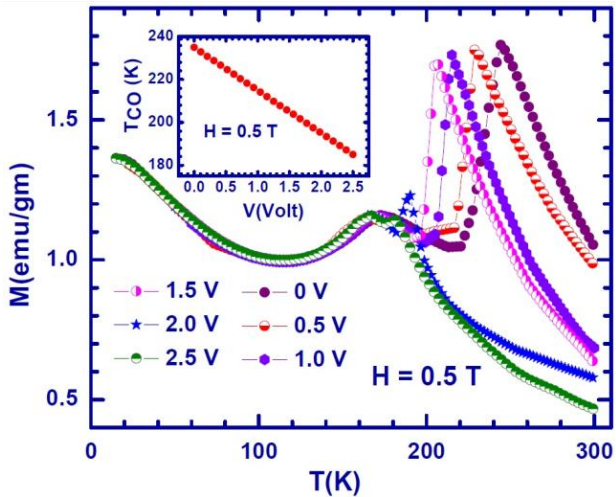

FIG. 2. Field-Cooled (FC) magnetization as a function of temperature (T) in an applied magnetic field of 0.5 T with applied voltage of 0 V, 0.5 V, 1 V, 1.5 V, 2 V and 2.5 V.

Even though the applied voltage affects the charge ordering, it has no obvious effect on the other two AFM transitions and the corresponding magnetization values. This is in dire contrast with the effect of the application of magnetic field on the ordering temperatures, as shown in Fig. 3, where we have shown the field-cooled magnetization of the same PCMO single crystal as a function of temperature at different applied magnetic fields starting from 1 T to 9 T.

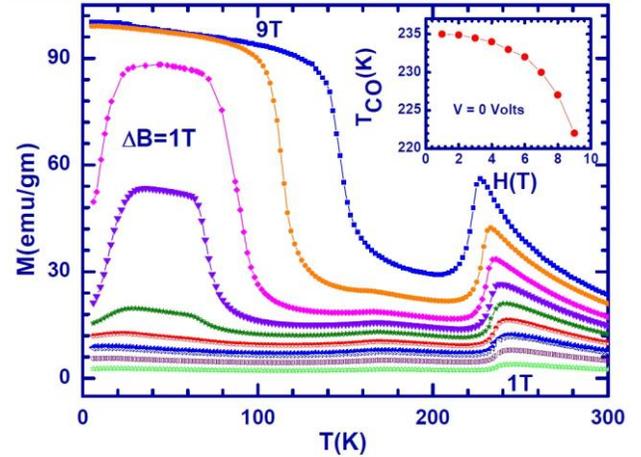

FIG. 3. Field-Cooled (FC) magnetization of PCMO as a function of temperature (T) in applied magnetic fields starting from 1 T to 9 T with ΔB = 1 T.

It is observed that the two antiferromagnetic transitions at $T_{N1}$ ~ 175 K and $T_{N2}$ ~ 25 K can be altered (melted) by applying large magnetic fields while the charge-ordering transition ($T_{CO}$ = ~235 K) cannot be destroyed even after applying a magnetic field as high as 9 T even though the $T_{CO}$ deceases (non linearly) with increasing applied magnetic fields (decreases only by 10% for 9 T, see inset of Fig. 3). The observed shift in charge-ordering transition temperature with magnetic field is in agreement with the observation by N. Biskup et. al. [10], where they could observe the vanishing of the charge ordered state only after the application of a magnetic field of ~11 T.

The change in magnetization as a function of applied voltage across the CO region is further confirmed through the magnetization measurements as a function of applied field at three different temperatures, 200 K, 225 K and 250 K. Figure 4 shows the magnetization as a function of magnetic field with different applied voltages of 0 V, 1 V and 2 V, respectively.

At 250 K, close to charge-ordering transition ($T_{CO}$), the change in magnetization as a function of applied voltage is very prominent; magnetization decreases as the applied voltage increases, as expected from the magnetization behaviour shown in Fig. 2. At 200 K, magnetization with applied voltages of 0 V and 1 V are almost same but a decrease in magnetization is observed at 2 V. However, at 225 K magnetization increases with applied voltage of 1 V in comparison with zero applied voltage but a further increase in voltage results in a decrease in the magnetic moment. The changes observed in magnetization as a function of magnetic field (in Fig. 4) at different temperatures with varied applied voltage are consistent

a)Email: himsharma@iitb.ac.in  b)Email: ashwin@ee.iitb.ac.in
c)Email: tomy@iitb.ac.in

with the results observed in the magnetization as a function of temperature (in Fig. 2).

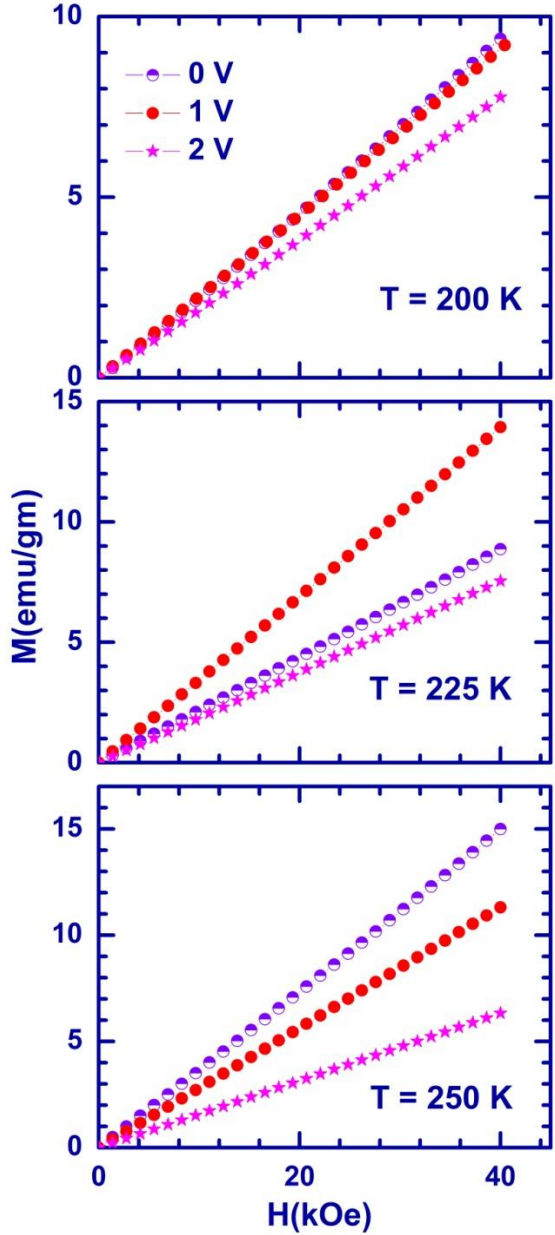

FIG. 4. Magnetization (M) curves of PCMO at 200 K, 225 K and 250 K measured with applied voltages of 0 V, 1 V and 2 V.

It is well known that $Pr_{0.6}Ca_{0.4}MnO_3$ (PCMO) exhibits magnetic field-induced metamagnetic[10,22,23] transition to a ferromagnetic state (FIFM) which persist up to a temperature T ≤ TCO. The manifestation of metamagnetic transition is a clear signature of the coexistence of the CO-antiferromagnetic (CO-AFM) phases and the ferromagnetic (FM) phases[10,22,23]. It will be now quite interesting to investigate whether the application of an electric field has any effect on these metamagnetic transitions. In order to confirm the occurrence of metamagnetic transitions in our crystal, we first measured the magnetization as a function of magnetic field at one particular temperature (15 K) with zero applied voltage, as shown in Fig. 5. The arrows denote the direction of the magnetic field sweep (0 T → + 7 T → 0 T → − 7 T → + 7 T). The metamagnetic transition is clearly visible (H = ∼ 6 T). Also, upon reversing the magnetic field, magnetization traces a completely different path. Hence the irreversibility in AFM to FM phase change shows a spin memory effect.

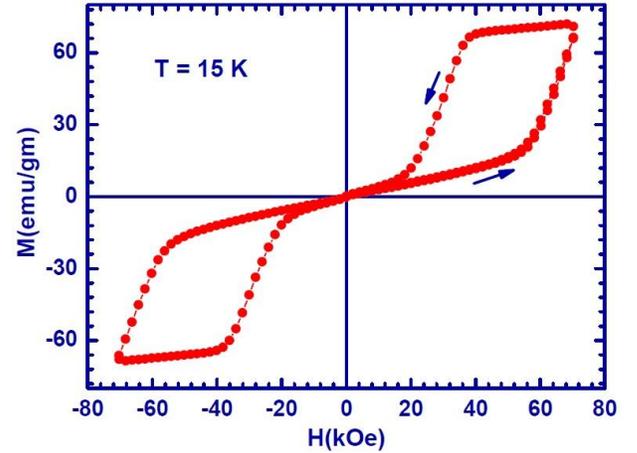

FIG. 5. Magnetization as a function of magnetic field at 15 K with zero applied voltage.

Now, in order to investigate the electric field effect on the metamagnetic transition as mentioned above, we have measured magnetization as a function of magnetic field (for one quadrant) at 15 K with different applied voltages. Figure 6 shows the variation of magnetization for applied voltages of 0.2 V, 0.5 V, 1.0 V and 1.5 V along with that of 0 V for comparison. It is very interesting to note that even a small voltage as low as 0.2 V is enough to affect the metamagnetic transition. As we increase the voltage, the metamagnetic transition shifts towards higher magnetic fields. For voltages ≥ 1.0 V, we could observe only the reversible, paramagnetic part, and not any traces of the metamagnetic transition since these transitions might be occurring at fields higher than our measurable field limit (7.0 T).

The results in this chapter in conjunction with N. Biskup et. al.[10], bring in motivating aspects regarding the melting of charge-ordered state and metamagnetic transitions in PCMO. Even with a very small applied voltage of 2.5 V, the charge ordered state can be completely suppressed which otherwise needs a large magnetic field, as large as 11 T. Thus the applied voltage can be considered equivalent to applied fields and a correlation can be brought in as shown in Fig. 7. The reduction in $T_{CO}$ with applied voltage

a) Email: himsharma@iitb.ac.in   b) Email: ashwin@ee.iitb.ac.in
c) Email: tomy@iitb.ac.in

is almost linear, whereas the shift of $T_{CO}$ is nonlinear when the magnetic field is used.

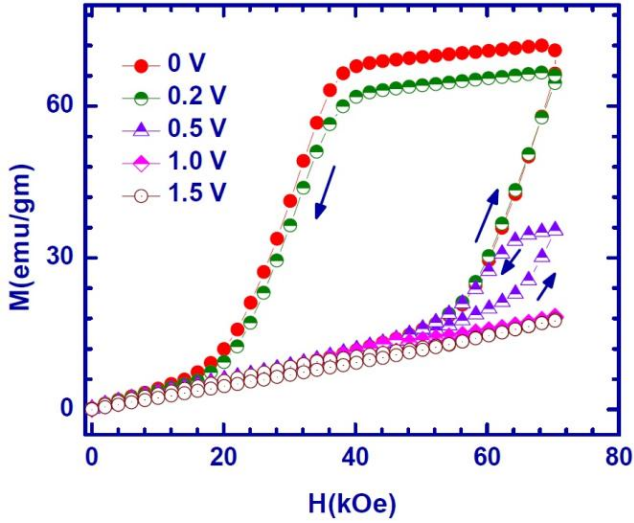

FIG. 6. Magnetization as a function of magnetic field at 15 K with applied voltages of 0.2 V, 0.5 V, 1.0 V and 1.5 V along with that of 0 V for comparison.

Another interesting observation is that the melting of CO can be achieved without affecting the other magnetic transitions by the application of the voltage (see Fig. 2). Whereas with the application of the magnetic field, the magnetic states below the charge ordering are completely transformed into new magnetic states (see Fig. 3). Even though the applied voltage acts equivalent to the magnetic field in suppressing the charge ordering, the same applied voltage has a contradictory role on the magnetization in the temperature range from 300 K down to the charge ordering temperature, where we observe a decrease in magnetization with the increase in applied voltage.

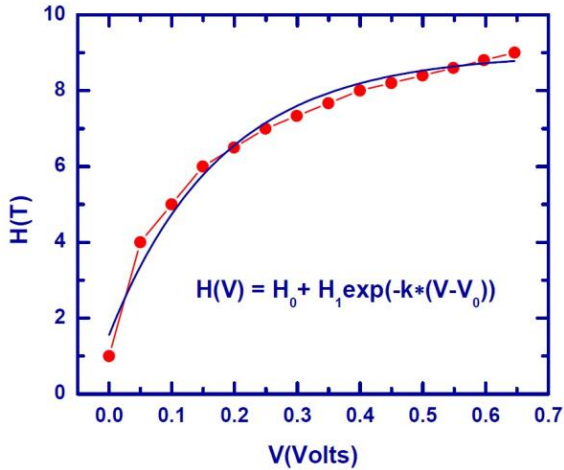

FIG. 7. Voltage as a function of applied magnetic field and blue line shows the fitting of the curve.

The exact reason why the magnetization in this region should decrease with the application of voltage is not clear, but may be explained on the basis of reports, where the fluctuation in the charge-ordered states is seen to persist even at temperatures above the room temperature[3,13,21]. Even though there is no clear understanding as to the reason for the melting of chargeordered state, one of the possibilities is the formation of Zener-polarons in PCMO reported by Daoud-Aladine et. al[13]. They observed that the Mn ions remain in an intermediate valence state due to the formation of Mn-Mn dimers, known as Zener-polarons (ZP), which has also been reported by J.-S Zhou, et. al.[21], for $La_{1-x}Sr_xMnO_3$ system previously. An applied electric field is expected to de-pin the randomly pinned charge carriers within the Mn pairs due to double exchange (DE) and a polaronic-like distortion[13]. The reason for the suppression of metamagnetic states by applied voltages needs further investigations. Low-power spintronics devices (e.g., spintronics field effect transistor) can be fabricated by using these materials as channel material of a prototypical field effect device. Also, it will be interesting to see whether such change in magnetization can also be observed directly by measuring the magnetization of PCMO thin film in the presence of applied gate voltage using insulating or ferroelectric gate[24].

In conclusion, we have shown that the magnetization, charge-ordered state and metamagnetic transitions can be tuned by applied electric fields in a PCMO single crystal. We have observed that the charge-ordered state can be completely melted by applying a few volts, which may be attributed to the de-pinning of randomly pinned charge carriers within the Mn pairs due to vibronic electronic states. Low-power spintronics devices (e.g., spintronics field effect transistor) can be fabricated by using these materials as channel material of prototypical field effect devices.

We are grateful for availability of the Institute central facility (SQUID-VSM) in the Department of Physics and Institute facility (MPMS-XL) in the Department of Chemistry, Indian Institute of Technology Bombay.

[a)]Email: himsharma@iitb.ac.in  [b)]Email: ashwin@ee.iitb.ac.in
[c)]Email: tomy@iitb.ac.in

[a]Email: himsharma@iitb.ac.in    [b]Email: ashwin@ee.iitb.ac.in
[c]Email: tomy@iitb.ac.in